\documentstyle[12pt,aasms4]{article}

\newcommand {\eg} {{\it e.g.}}
\newcommand {\ea} {{\it et~al.}}
\newcommand {\be} {\begin{equation}}
\newcommand {\ee} {\end{equation}}

\begin{document}

\title{X--ray Observations of BL Lacertae During 1997 Outburst and its
Association with Quasar-like Characteristics}
\author{Greg M. Madejski$^{1,2}$, Marek Sikora$^{3}$, Tess Jaffe$^{1,4}$, Micha{\l } B{\l }a\.zejowski$^{3}$, Keith Jahoda$^{1}$, \& Rafa{\l} Moderski$^{3,5}$} 
\affil{$^1$Lab for High Energy Astrophysics, NASA/Goddard, Greenbelt, MD
20771 \\
$^2$also with the Dept. of Astronomy, Univ. of Maryland, College Park; \\
$^3$Copernicus Astronomical Center, Bartycka 18, 00-716 Warsaw, Poland; \\
$^4$also with Raytheon STX; \\
$^5$JILA, University of Colorado, Campus Box 440, Boulder, CO 80309}

\lefthead{Madejski \ea}
\righthead{X-ray Observations of BL Lac}

\begin{abstract}
  This paper reports on the X--ray emission from BL Lacertae during its
July 1997 outburst as observed with the Rossi X--ray Timing Explorer
(RXTE), compares the RXTE data to previous measurements, and interprets the
overall electromagnetic emission in the context of the currently popular
theoretical models.  The source is bright and variable, with the 2 -- 10
keV flux approximately two to three and a half times higher than measured
in November 1995 by {\sl Asca}.  The spectrum is also harder, with power
law energy indices of $\sim 0.4 - 0.6$, compared to $\sim 0.9$ in
Nov. 1995.  Both in the optical band, where BL Lacertae now shows broad
emission lines, and in the X--ray band, where the spectrum is hard, the
overall electromagnetic distribution of BL Lacertae is similar to that
observed in blazars associated with quasars rather than to that seen in the
more common High-energy - peaked BL Lac - type objects (HBLs).  We argue
that the high energy (X--ray and $\gamma$--ray) emission from BL Lacertae
consists of two spectral components: X--rays are produced by Comptonization
of synchrotron radiation, while the $\gamma$--rays produced by
Comptonization of the broad emission line flux.
\end{abstract}

\keywords{X--rays:  observations -- galaxies:  active -- 
quasars:  BL Lacertae -- theory:  radiation mechanisms}

\section{INTRODUCTION}

The eponymous BL Lacertae has been identified as a counterpart of the
variable radio source VRO 42.22.01 by Schmitt (1968); subsequent work by
Oke, Neugerbauer, \& Becklin (1969) as well as DuPuy \ea  (1969)
revealed a featureless spectrum, devoid of emission or absorption lines.
In 1972, Strittmatter \ea (1972) suggested that BL Lac-type objects are
akin to Quasi-Stellar Radio Sources, but distinguished by the absence of
emission lines.  Detailed spectroscopy of BL Lacertae by Miller \& Hawley
(1977) revealed an absorption line system at a redshift of 0.069,
presumably due to the galaxy hosting BL Lacertae.  At the 1978 conference
devoted to BL Lac - type objects, it was suggested that the presence or
absence of optical and/or UV emission lines is perhaps less relevant to the
physical structure of blazars than their rapid variability, the existence
of compact radio sources, and a large degree of polarization (cf. numerous
articles in Wolfe 1978).  The term ``blazar,'' dating back to that
conference, includes both ``lineless'' BL Lac objects as well as ``lined''
quasars showing the same characteristics as BL Lac objects, plus emission
lines.  Independently, further observations of compact radio sources,
particularly of their variability and of the detection of superluminal
expansion, led to the suggestion that at least the matter responsible for
the radio emission is moving at a relativistic speed at an angle close to
the line of sight, perhaps in a jet-like structure (see, \eg, Blandford \&
Rees 1978).

The most convincing evidence of similarity between the two sub-classes of
blazars is the fact that both often show show strong and variable GeV
$\gamma$--ray emission (cf.  von Montigny \ea 1995).  Of all classes of
extragalactic sources, only blazars are strong GeV $\gamma$--ray emitters.
In contrast, ordinary radio-quiet quasars and Seyfert galaxies, regardless
of their optical, UV, and X--ray fluxes, are not detected above $\sim 1$
MeV.  In general, blazars show two distinct peaks in their $E \times F(E)$
spectra, with one located in the infrared - through - X--ray band, and
another located in the MeV to GeV (or even TeV) $\gamma$--rays (cf. Von
Montigny \ea 1995).  (In some cases, other, narrower peaks are present, but
these are generally attributed to isotropic emission due to the host galaxy
or the accretion disk.)  The rapid variability observed in the
$\gamma$--rays implies that the $\gamma$--ray source must be very compact.
The simplest way to avoid an excessive opacity to $\gamma - \gamma$ pair
production is to invoke relativistic boosting of $\gamma$--ray emission;
most likely, the entire continuum emission in blazars is Doppler-boosted
via a jet-like structure.  BL Lacertae itself indeed shows GeV emission in
the EGRET observations; in fact, the $\gamma$--ray flux ($E >$ 100 MeV)
increased from an upper limit of $30 \times 10^{-8}$ photons cm$^{-2}$
s$^{-1}$ in Oct. 1994 to a detection of $40 \pm 12 \times 10^{-8}$ photons
cm$^{-2}$ s$^{-1}$ in Jan. 1995 (cf. Catanese \ea 1997).

X--ray surveys conducted over the last 20 or so years with satellites such
as HEAO-A, Einstein Observatory, ROSAT, and {\sl Asca}, revealed that BL
Lac objects are generally strong X--ray emitters, with the X--ray spectrum
generally well-described as a power law. (For a recent review, see, \eg,
Kubo \ea 1998.)  Those surveys revealed that the X--ray emission from BL
Lac objects obeys a peculiar correlation: the ratio of the X--ray to
optical fluxes was greater for objects where the ratio of the
radio-to-optical fluxes was smaller (cf. Maraschi 1988).  This correlation
suggested that instead of dividing blazars by the presence/absence of
emission lines, a better classification would rely on the overall spectrum.
In this context, the BL Lac objects with the low energy peak located in the
UV or X--rays -- usually found via X--ray surveys -- would be labeled as
``High-energy peaked BL Lacs'' or HBLs (cf.  Giommi, Ansari, \& Micol
1995), while those with the lower energy peak in the IR were labelled as
``Low-energy peaked BL Lacs'' or LBLs.  The LBLs show broad-band spectra
similar to blazars associated with lined, compact, flat radio-spectrum
quasars (cf. Sambruna, Maraschi, \& Urry 1986), which we will call here
``quasar-hosted blazars,'' or QHBs.

In this classification BL Lacertae is an LBL.  Its spectrum peaks at $\sim
10^{14}$ Hz (cf. Kawai \ea 1991), and, historically, the optical spectrum
was devoid of any emission lines.  This changed within a few years before
May 1995.  A serendipitous observation at the Hale Observatories, intended
for calibration purposes (Vermeulen \ea 1995) revealed emission lines with
equivalent H$\alpha$ width of $\sim 7$ \AA \ or more and FWHM of $\sim
4000$ km s$^{-1}$.  Corbett \ea (1996) confirmed the presence of the lines
and inferred that the line flux increased four-fold.  BL Lacertae,
therefore, lost its defining characteristics.  Interestingly enough, this
happened around the epoch of the increase of the GeV flux.

Following reports that BL Lacertae entered an active, high state in June
1997 (Noble \ea 1997), an impromptu monitoring campaign was organized.  We
observed it with the Rossi X--Ray Timing Explorer (RXTE) (Madejski, Jaffe,
\& Sikora 1997).  The questions to be addressed were: what are the changes
in the X--ray emission associated with the emergence of emission lines and
the occurrence of the flare, and what constraints can the broad-band data
impose on models of blazar jets?  We report on X--ray observations (RXTE,
as well as previous X--ray observations) in Sec. 2.  In Sec. 3 and 4, we
put the X--ray observations in the context of the broad-band spectrum and
discuss the most plausible models for the radiative processes in the
source, and in Sec. 5 we summarize our results.

\section{OBSERVATIONS}

Prior to the RXTE observation in June 1997 -- which we describe in Sec. 2.3
-- BL Lacertae was observed with the ROSAT PSPC and by {\sl Asca}.  We
extracted these data from the HEASARC archives, and found that these show
softer continuum and lower flux than the RXTE data.  In addition, the
well-exposed {\sl Asca} data are very helpful in determining the low energy
(photoelectric) absorption, which is probably not related to the source,
and is assumed to be non-variable.

\subsection{Asca Observations and Spectral Fitting}

{\sl Asca} observed BL Lacertae on 1995 November 22 for approximately 30
ks.  The {\sl Asca} data were screened using the {\tt ftool ascascreen} and
the standard screening criteria.  The pulse-height data for the source were
extracted using spatial regions with a diameter of $3'$ (for SISs) and $4'$
(for GISs) centered on the nominal position of BL Lacertae, while
background was extracted from source-free regions of comparable size away
from the source.  The PHA data were subsequently rebinned to provide at
least 20 counts per spectral bin.  For the SIS data, we used the response
matrix as appropriate for the observation epoch, generated via the {\tt
ftool sisrmg} v. 1.1; for the GIS data, we used the nominal (v. 4.0)
response matrices.  For both instruments, we used the telescope effective
areas via the {\tt ftool ascaarf} v. 2.72.  The details of the observation
(including the net counting rates) are given in Table 1.

We fitted the full-band PHA data using a simple power law absorbed at the
low energies by neutral gas with Solar composition and cross-sections as
given by Morrison \& McCammon (1983); the results of the fits are given in
Table 1.  Using the data for all four {\sl Asca} detectors over their full
bandpass (0.6 -- 10 keV), we get the following best fit (yielding $\chi^2$
of 737 for 778 PHA channels): energy power law index $\alpha = 0.94 \pm
0.04$, and equivalent hydrogen column density of absorbing gas $N_{\rm H,
X-ray} = 2.7 \pm 0.2 \times 10^{21}$ cm$^{-2}$ (all errors are $90\%$
confidence regions, meaning that they are determined from the values of
fitted parameters at $\chi^2_{\rm min} + 2.7$).

While this fit is statistically acceptable, the value of the absorbing
column in this simple absorbed power law model may be inconsistent with the
Galactic value.  Since BL Lacertae is located relatively close to the
Galactic plane, it is necessary to consider the possible contribution from
the Galactic molecular gas in addition to that associated with the usual 21
cm atomic hydrogen measurement.  In the case of BL Lacertae, this was
detected in emission in $^{12}$CO (Kazes \& Crovisier 1981; Bania,
Marscher, \& Barvainis 1991) and in $^{13}$CO (Crovisier, Kazes, \& Brillet
1984), as well as in absorption in $^{12}$CO (Marscher, Bania, \& Wang
1991).  The absorbing column consists therefore of two components: that
associated with neutral hydrogen, inferred from the 21 cm measurements of
Dickey \ea (1983) of $N_{\rm H, 21cm}$ of $1.8 \times 10^{21}$ cm$^{-2}$,
and the molecular component.  Bania \ea (1991) measure an integrated CO
emission $W_{\rm CO}$ of 4.6 K km s$^{-1}$, adopt the ratio $N_{\rm H,
mol}$ / $W_{CO}$ of $6 \times 10^{20}$ K$^{-1}$ km$^{-1}$ s cm$^{-2}$, and
infer that the column of the molecular component corresponds to $N_{\rm H,
mol}$ of $\sim 2.8 \times 10^{21}$ cm$^{-2}$, yielding the total column
$N_{\rm H, tot}$ of $\sim 4.6 \times 10^{21}$ cm$^{-2}$, significantly
larger than $N_{\rm H, X-ray}$.

There are two possible reasons for this discrepancy.  Regarding the CO
measurements, the conversion of $W_{CO}$ to $N_{\rm H, mol}$ may be
unreliable (and direction-dependent); de Vries, Heithausen, \& Thaddeus
(1987) suggest that towards Ursa Major, the conversion of $N_{\rm H, mol}$
/ $W_{CO}$ of $1 \pm 0.6 \times 10^{20}$ K$^{-1}$ km$^{-1}$ s cm$^{-2}$ may
be more appropriate.  When applied to the case of BL Lacertae, this would
yield $N_{\rm H, mol}$ of $\sim 0.5 \times 10^{21}$ cm$^{-2}$, and $N_{\rm
H, tot}$ of $\sim 2.3 \times 10^{21}$ cm$^{-2}$, which is now less than
$N_{\rm H, X-ray}$.  If the value of $N_{\rm H, 21cm}$ is indeed accurate,
one could obtain an agreement between $N_{\rm H, X-ray}$ and $N_{\rm H,
tot}$ using the conversion $N_{\rm H, mol}$ / $W_{CO}$ of $\sim 2
\times 10^{20}$ K$^{-1}$ km$^{-1}$ s cm$^{-2}$, which is probably
acceptable.  

Alternatively, if the $N_{\rm H, tot}$ of $\sim 4.6 \times 10^{21}$ derived
by Bania \ea (1991) is indeed correct, this would imply that our simple
power law model for the X--ray spectrum emitted by BL Lacertae is
incorrect.  This would then require that the intrinsic spectrum hardens
towards higher energies.  Such a gradually hardening spectrum is entirely
possible, given the fact that the 1997 RXTE observation -- with the
bandpass extending beyond {\sl Asca's} -- implies an even harder index than
that in the simple power law model applied to the {\sl Asca} data.  As an
alternative, we thus adopt a model consisting of a sum of two power laws,
both absorbed by the fixed column suggested by Bania \ea (1991) of $4.6
\times 10^{21}$ cm$^{-2}$. (We note here that a commonly used broken power
law model for a description of such a hardening (concave) spectrum is
unphysical, and we do not consider it here.)  In this case, we obtain the
``soft'' power law index (dominating below $\sim 1$ keV) of $3.4 \pm 0.7$
and the ``hard'' index (dominating above $\sim 1$ keV) of
$0.88^{+0.09}_{-0.14}$.  This yields $\chi^2$ of 736 for 778 PHA channels.
We conclude that we cannot distinguish between the two models purely on the
statistical basis.  Nonetheless, the double power law model, if correct,
may be attractive, suggesting that the {\sl Asca} data reveal
simultaneously the soft component -- presumably the ``tail'' of the
synchrotron emission -- and the hard component, presumably the onset of the
Compton component.  However, we stress that this is {\sl not} a unique
representation of the {\sl Asca} data, and more precise X--ray observations
(with an instrument of better spectral resolution) are required to
precisely measure the absorbing column via the measurement of the
individual edges.  In any case, the observed 2 -- 10 keV flux of BL
Lacertae during the 1995 November observation for either of the above
models is $\sim 9 \times 10^{-12}$ erg cm$^{-2}$ s$^{-1}$, with a 10\%
nominal error.  In the 0.5 -- 2 keV band (useful for a comparison with the
ROSAT data), it is $\sim 3.6 \times 10^{-12}$ erg cm$^{-2}$ s$^{-1}$, again
with a 10\% nominal error.

\subsection{ROSAT PSPC Observations and Spectral Fitting}

ROSAT observed BL Lacertae with the PSPC starting on December 22, 1992, in
2 short pointings about two days apart.  The data were reduced using
standard ROSAT PSPC procedures via the {\tt ftool xselect} v. 1.4.  The
source PHA file was extracted from the event data using a circular
extraction region of $3'$ in diameter, while the background was extracted
from an annular region with the inner and outer diameters of $10'$ and
$20'$, respectively.  The data were fitted using the detector response
matrix {\tt pspcb\_gain2\_256.rmf} and telescope effective area file
prepared using the {\tt ftool pcarf} v. 2.1.1.  The resulting counting rate
was $0.16 \pm 0.01$ count s$^{-1}$ with no measurable variability apparent
in the data.

A 2 ks observation of a relatively faint source with strong absorption such
as BL Lacertae with an instrument with a relatively soft X--ray response
such as the ROSAT PSPC yields data with only moderate statistical quality.
Nonetheless, an interesting conclusion can be drawn by comparing these data
against the {\sl Asca} and the 21 cm / $^{12}$CO data sets.  An absorbed
power law model of the form as above yields an acceptable fit ($\chi^2$ is
30.4 for 40 PHA bins) with $\alpha = 4.1 (-1.6, +2.4)$ and $N_H = 7 \pm 3
\times 10^{21}$ cm$^{-2}$; however, this best-fit value of the absorbing
column is significantly higher than that inferred from the 21 cm /
$^{12}$CO or the {\sl Asca} data.  The most likely explanation is that the
underlying continuum is more complex, curving downward.  Independently of
the severe conflict with the absorption measured by other instruments, such
a steep power law cannot extend to lower energies indefinitely, and thus it
must have a convex shape.  Instead, such a shape may be approximated as a
broken power law (which {\sl can} be a physically realistic approximation
of a spectrum; see, \eg, the comment after Eq. 20), or an exponentially
cutoff power law.  Again, the quality of the data is modest, but to shed
some light on the possible shape of the X--ray spectrum of BL Lacertae
above $\sim 1$ keV in a very low state, we assumed that the absorbing
column is indeed $2.7 \times 10^{21}$ cm$^{-2}$ and that the index below
the break is the same as observed in {\sl Asca}, $\alpha_{lo} = 0.9$.  With
this, we infer that for a broken power law model, the break energy $E_b$ is
$1.0 < E_b < 1.6$ keV, and the index above the break is $\alpha_{hi} > 2.2$
($\chi^2$ is 29.3).  For an exponentially cutoff power law model with an
underlying index $\alpha = 0.9$, we infer the e-folding energy $E_c$ to be
$1.0 < E_c < 2.3 $ keV, with $\chi^2$ of 34.8.  Therefore a broken power
law yields a better fit.  The fits are summarized in Table 1; regardless of
the model, the 0.5 -- 2 keV flux is $\sim 1.5 \times 10^{-12}$ erg
cm$^{-2}$ s$^{-1}$, with a nominal 10\% error; this is roughly a half of
the {\sl Asca} flux in the same bandpass.  In summary, when BL Lacertae was
relatively faint, its intrinsic spectrum showed convex curvature and is
well described by a phenomenological model of a broken power law.

\subsection{RXTE Observations and Spectral Fitting}

RXTE observations consisted of seven pointings over six days listed in
Table 2; each pointing covered one or two orbits, interrupted only by the
Earth occultations.  These observations were scheduled during
low-background orbits, i.e. those relatively free of passages through the
South Atlantic Anomaly (SAA).  The PCA instrument (Jahoda \ea 1996)
consists of 5 individual passively collimated, co-aligned, gas-filled
proportional counter X--ray detectors, sensitive over the bandpass of 2 --
60 keV, each having an open area of $\sim$ 1300 cm$^{-2}$.  The field of
view of the instrument has roughly a triangular response, with FWHM of
$\sim 1^{\rm o}$.  The HEXTE instrument (Rothschild \ea 1998) consists of
two detector clusters, each having four NaI / CsI scintillation counters,
sensitive over the bandpass of 15 -- 250 keV, with the total effective area
of $\sim$ 1600 cm$^{2}$, and a field of view also of $\sim 1^{\rm o} \times
1^{\rm o}$.

For the PCA data, only three or four of the five detectors were operating.
For maximum consistency among all observations, we used only the three
detectors (known as PCU 0, 1, and 2) which were turned on for all the
observations.  The selection criteria were: the elevation angle over the
Earth limb greater than $10^{\rm o}$, and pointing direction less than
$0.02^{\rm o}$ away from the nominal position of the source at RA(2000) $=
22^{\rm h} 02^{\rm m} 43^{\rm s}.2$, Dec(2000) = $42^{\rm o} 16^{'}
40^{''}$.

\subsubsection{PCA data}

For a source as faint as BL Lacertae, more than half of the counts
collected in the PCA are due to unrejected instrumental and cosmic X--ray
background.  To maximize the signal-to-noise ratio, we used only the top
layer PCA data from the ``Standard 2'' mode.

The PCA background, instrumental plus cosmic, has been modeled from
observations of blank (i.e. not near known X--ray sources) high latitude
($|b| > 30^{\rm o}$) sky.  The raw counting rate varies with satellite
latitude (i.e.  with a period about half of the 96 minute orbital period)
and with activation induced by the South Atlantic Anomaly (SAA).  At least
3 time constants are present in the unrejected background ($\sim$ 20
minutes, $\sim$ 4 hours, and $\sim$ 4 days).  The latitude variation is
primarily due to the instantaneous particle environment while the
activation is primarily due to the recent history of passages through the
SAA.  The background is parameterized by the so-called "L7" rate, which is
derived from the two-fold coincidence Lower Level Discriminator (LLD) event
rates present in the Standard 2 data (Jahoda \ea 1996).  In particular,
``L7'' consists of the instantaneous sum of signals derived from coincident
``events'' on anodes L1+R1, L2+R2, L1+L2, R1+R2, R3+L3, R2+R3, L2+L3 (Zhang
\ea 1993.)  This rate tracks the particle-induced background as well as the
short and long time constants.  To this model, a time dependent term is
added.  The rate measured by the HEXTE particle monitor (the only detector
onboard RXTE which operates during the SAA passages) is integrated through
each SAA pass, and a term proportional to the sum of recent SAA rates
$\times e^{-(t-t_{saa}/\tau)}$ is included.  The spectral shape of this
activation term is assumed to be constant, and the amplitude is determined
by comparison of the observed total background in orbits just following SAA
passages with the observed background in orbits far from SAA passages.  The
distribution of residuals to a background-subtracted count rate from a
single blank sky region suggests a residual 1 $\sigma$ systematic error of
0.15 count s$^{-1}$ (3 PCUs, 2 - 10 keV) (see Jahoda \ea 1999, in
preparation).  The resulting background-subtracted count rates for the 7
observations are shown in Table 2 and Fig. 1.  Visual inspection indicates
that the source varies significantly, and that the variability is
undersampled; we can only state that the variability time scale is a day or
less.

Since the data for the background estimation were collected from blank sky
observations, the average Cosmic X--ray Background (CXB) is included in the
background estimate, by construction.  The contribution of the CXB flux to
the total observed flux in the PCA data can be scaled from the measurement
given by Marshall \ea (1980) of 3.2 keV cm$^{-2}$ s$^{-1}$ sr$^{-1}$
keV$^{-1}$ at 10 keV, with a spectral shape well-approximated as a thermal
bremsstrahlung with $kT = 40$ keV.  Since the solid angle of the PCA
collimator is $\sim 3.2 \times 10^{-4}$ sr (Jahoda \ea 1996), the 2 - 10
keV flux is $1.7 \times 10^{-11}$ erg cm$^{-2}$ s$^{-1}$.

While it is possible to estimate the contribution of the {\sl mean} CXB
flux, the CXB is not uniform from all directions in the sky.  The
fluctuations in the CXB on the solid angle scale of the PCA collimator can
be estimated by scaling the HEAO-1 A2 fluctuations by the ratio of square
root of the solid angles of A2 and PCA detectors (Shafer 1983; Mushotzky
\& Jahoda 1992, Fig. 1, with the correction that the mean counting
rate should be 3.5 count s$^{-1}$, and not 5.6 count s$^{-1}$ as stated in
the caption).  The 1 $\sigma$ CXB fluctuation is about $10\%$ of the CXB
contribution, or $\sim$ 0.2 count s$^{-1}$ per PCA detector (2 -- 10 keV).
While the variability can be reliably measured at much smaller levels, this
represents a limiting uncertainty in the determination of the absolute flux
for observations using the PCA alone.

\subsubsection{HEXTE data}

The HEXTE background is continuously measured, as each cluster alternately
rocks to one of two off-source pointings (on alternate 16 second intervals
for this observation).  This gives an effectively simultaneous background
measurement at four positions around the source.  This background is
dominated by internal activation (Gruber \ea 1996).  A significant
instrument deadtime (15 -- 40\%), however, introduces some uncertainty in
the absolute flux level.  This deadtime is largely particle-induced and is
therefore significant even for faint sources.  A deadtime correction factor
is calculated from the charged particle rates to bring the exposure to
within a few percent of its actual value (based on 16 s timescale
observations of the Crab).  The resulting residuals for the background
estimation using this procedure are 1\% of background or less (Rothschild
\ea 1998).  However, while BL Lacertae was detected with HEXTE at $\sim
0.5$ net count s$^{-1}$, each data segment had too low a signal-to-noise
ratio to measure the spectrum, and we could only do so for the summed data
(cf. Table 2).

\subsubsection{Spectral fitting of the RXTE data}

We fit the PHA spectrum from each observation to a model including a simple
power law which is photoelectrically absorbed at low energies by neutral
gas with Solar abundances and with cross-sections given by Morrison \&
McCammon (1983) with a fixed column density of $2.7 \times 10^{21}$
cm$^{-2}$ as determined from the {\sl Asca} observations.  (We note that
adopting a column of $4.6 \times 10^{21}$ cm$^{-2}$ does not change our
conclusions significantly.)  In our fits, we used the instrumental response
matrix generated via {\tt ftool pcarmf} v. 3.5, which corrects for the
slight PCA gain drift ($\sim$ 1\% over 2 years) and energy -- to -- channel
conversion table {\tt pca\_e2c\_e03v04.fits}, as appropriate for the
observation epoch of BL Lacertae.  The results in Table 2 indicate that the
data for each pointing are well-described by the absorbed power law model,
but we do observe spectral variability from one pointing to another, with
the energy power law index $\alpha$ varying from $0.42\pm0.08$ to
$0.80\pm0.06$, with no clear correlation of the index with the flux level.
To determine the average spectral shape to the highest possible energies
(where additional information is gained from the HEXTE data), we also
co-added all data.

For simultaneous analysis of PCA and HEXTE data, there is some uncertainty
in the relative normalization which must be taken into account.  This
difference is not determined uniquely as yet, but in general, the PCA
agrees more closely with previous data (namely OSSE and {\sl Ginga}) and is
therefore taken as the baseline against which the HEXTE data are
normalized.  This factor is generally $\sim 0.7$ for the fits using the
effective areas released in {\tt ftools} v. 4.1.  The addition of the HEXTE
data does not change the resulting spectral fit parameters, but implies
that the hard X--ray spectrum observed in PCA extends to higher energies
($\sim 70$ keV); the summed PCA and HEXTE spectra are plotted in Fig. 2.
The best-fit spectral model gives a mean 2 -- 10 keV flux of $2.6 \pm 0.3
\times 10^{-11}$ erg cm$^{-2}$ s$^{-1}$.  This value has an additional
associated uncertainty of $\sim 0.2 \times 10^{-11}$ erg cm$^{-2}$ s$^{-1}$
from the fluctuations of the CXB.

\section{DISCUSSION}

The RXTE observations of BL Lacertae conducted in July 1997 show that the
source was bright and variable in the X--ray band, with the X--ray spectrum
significantly harder than observed during the periods of lower brightness
and activity; the mean 2 -- 10 keV flux was $2.6 \times 10^{-11}$ erg
cm$^{-2}$ s$^{-1}$, about 3 times greater than that observed by {\sl Asca}
in November 1995.  A direct comparison with the ROSAT PSPC observation in
December 1992 is not possible, but the 0.5 -- 2 keV flux in 1992 was about
a half of that in Nov. 1995, so the mean July 1997 flux must have been at
least 6 times greater than in December 1992.  BL Lacertae was also observed
by {\sl Ginga} in 1988 (Kawai \ea 1991).  The June 15 1988 observation
implies a 2 -- 10 keV flux of $7.6 \times 10^{-12}$ erg cm$^{-2}$ s$^{-1}$
with $\alpha$ of $0.71 \pm 0.07$, while for July 17 1988 observation, the 2
-- 10 keV flux was $5.5 \times 10^{-12}$ erg cm$^{-2}$ s$^{-1}$ with
$\alpha$ of $1.16 \pm 0.24$.  In general, the X--ray spectrum of BL
Lacertae appears to be harder when the source is brighter.  This is
illustrated in Fig. 3, where we plot the unfolded summed PCA spectrum
together with the {\sl Asca} GIS and ROSAT PSPC spectra.

The analysis of the {\sl Asca} data by Kubo \ea (1998), as well as many
papers published earlier (cf. Sambruna \ea 1996; Worrall \& Wilkes 1990),
imply that HBL-type blazars generally have relatively soft X--ray ($\alpha
> 1$) spectra, while QHBs show harder spectra, with $\alpha < 1$.  The
spectra of LBLs, the class of objects where BL Lacertae belongs, are
intermediate.  For BL Lacertae, the shape of the spectrum is related to the
state of the source.  Interestingly, the X--ray spectral characteristics of
BL Lacertae appear to be ``HBL-like'' when the emission lines are weak or
absent, and ``QHB-like'' when the emission lines are strong; an intriguing
possibility (although by no means certain, depending on the details of the
Galactic absorption; cf. Section 2.1) is that during the intermediate state
of the source, shortly after the emergence of the emission lines, the
spectrum simultaneously consisted of both the LBL-like (hard) and HBL-like
(soft) components, with comparable fluxes at $\sim 1$ keV.

The observations of blazars imply that the entire continuum (including both
the low- and high-energy components mentioned in Sec. 1) arises in a
relativistic jet.  The polarization and the local power-law shape of the
spectrum of the low energy component suggest that the process responsible
for emission in this spectral region is synchrotron radiation by highly
relativistic electrons, accelerated by an as yet unknown mechanism.  For
the high energy peak, the best current model is Compton-upscattering of
soft photons by the same electrons that produce synchrotron radiation via
interaction with magnetic field.  The origin of those soft photons is under
debate: they can be either synchrotron photons internal to the jet, as in
synchrotron self-Compton models (cf. Blandford \& K\"onigl 1979; K\"onigl
1981; Ghisellini \& Maraschi 1989), or they can be external to the jet,
either from the accretion disk (Dermer, Schlickeiser, \& Mastichiadis
1992), or else from broad emission line clouds and/or intercloud material
(Sikora, Begelman, \& Rees 1994).  Perhaps the best current picture has the
former mechanism dominating the $\gamma$--ray emission in HBL-type blazars,
which usually do not show broad emission lines (but see, \eg, Padovani \ea
1998), while the latter dominates in QHBs, blazars associated with quasars
(cf. Madejski \ea 1997).  In this context, the soft X--ray spectra of HBLs
are the ``tails'' of the synchrotron component (and thus produced by the
most energetic electrons), while the hard spectra of QHBs are emitted via
the Compton process by the lower energy electrons.

The overall broad-band spectrum of BL Lacertae, including the July 1997
data, is plotted in Fig. 4.  The two peaks generally present in blazar
spectra are apparent in the plot.  However, the fact that the $\gamma$--ray
spectrum is hard -- with a spectral index which is lower (harder) than the
two-point spectral index between the hard X--rays and the beginning of the
EGRET spectrum -- and that it lies below the extrapolation of the X--ray
power-law spectrum, suggests that that the total high energy spectrum
consists of two separate components. Below, we follow the suggestion
presented by us at the November 1997 HEAD meeting that the high energy
spectrum of this source as measured in July 1997 actually consists of two
components, one radiated via the synchrotron self-Compton process,
dominating in the X--ray band, and another radiated via
external-radiation-Compton, dominating in the GeV $\gamma$--ray band, and
in the context of such a scenario, we estimate the physical parameters of
the radiating plasma.

\section{THEORETICAL MODELS}

In our study of the radiative processes operating in the jet of BL
Lacertae, we use the instantaneous spectrum averaged over the available
July 1997 data that are simultaneous with the EGRET observation (Fig. 4).
The optical data (Bloom \ea 1997) show that the July 1997 high state of BL
Lacertae is a superposition of many flares (and probably the same is true
for other spectral bands). We thus interpret these flares as a result of
the formation of relativistic shocks, which, within a given distance range
in a jet, effectively accelerate relativistic particles.  We assume that
the transverse size of these shocks is
\be
a \simeq c \Delta t \Gamma ,
\ee
where $\Delta t $ is the observed time scale of the flare 
and $\Gamma$ is the bulk Lorentz factor of the radiating matter.  

We investigate two models, the SSC (synchrotron-self-Compton) where the GeV
radiation results from Comptonization of the intrinsic synchrotron
radiation, and the ERC (external-radiation-Compton) where the GeV radiation
is produced by Comptonization of the broad emission line light.  The models
are specified by adopting the following parameters describing activity of
BL Lac in July 1997:

\vskip 0.1 cm\noindent
- location of the peak of the low-energy 
(synchrotron) component - $h\nu_S \sim 1 {\rm eV}$; 

\vskip 0.1 cm\noindent
- location of the peak of the high-energy (Compton) component - 
$h\nu_C \ge 10{\rm GeV}$;

\vskip 0.1 cm\noindent
- synchrotron luminosity - $L_S \sim 2 \times 10^{45}$ erg s$^{-1}$;

\vskip 0.1 cm\noindent
- Compton luminosity - $L_C \sim 8 \times 10^{45}$ erg s$^{-1}$;

\vskip 0.1 cm\noindent
- time scale of a  flare - $\Delta t \sim 8$ hrs;

\vskip 0.1 cm\noindent
- typical energies corresponding to broad emission line frequencies -
$h\nu_L \sim 10$ eV;

\vskip 0.1 cm\noindent
- energy spectral index in the X--ray band - $\alpha_X \sim 0.5$. 

\vskip 0.1 cm
To determine the parameters of the ERC model, we also need to know the
luminosity of the external radiation. We derive it by using the
measurements of the $H_{\alpha}$ line in June 1995 (Corbett \ea 1996) and
by assuming that the line intensity ratios in BL Lacertae are the same as
those in quasars.  Using the ``line-bolometric'' correction (Celotti,
Padovani \& Ghisellini 1997), we find $L_{BEL} \sim 4 \times 10^{42}$ erg
s$^{-1}$ and adopt it for the July 1997 flare.

\subsection{Synchrotron Self-Compton}

In the SSC model, $\nu_{C} \equiv \nu_{SSC}$, and we have
\be 
h\nu_{S} \simeq {\gamma_{m}'}^2 (B'/B_{cr})\Gamma \,{\rm m_ec}^2
\ee
and
\be
h\nu_{C} \simeq h\nu_{S} {\gamma_m'}^2,
\ee
where $B'$ is the intensity of the magnetic field, $B_{cr} \equiv 2\pi
m_e^2c^3/h e \simeq 4.4 \times 10^{13}$ Gauss, and $\gamma_m'$ is the
Lorentz factor at which the energy distribution of electrons has a high
energy break/cutoff. All primed quantities are measured in the comoving
frame of the flow in the active region.

Assuming that observer is located at an angle $\theta_{obs} \sim 1/\Gamma$
from the jet axis and noting that $L_S \sim \Gamma^4 L_S'$, we find that
the ratio of the SSC peak luminosity to the synchrotron luminosity is given
by (Sikora \ea 1994; Ghisellini, Maraschi, \& Dondi 1996)
\be
{L_C \over L_S} = 
{u_S' \over u_B'} \simeq
{L_S' \over 4\pi a^2 c} {8\pi \over {B'}^2} \simeq 
{2  L_S \over c^3 \Delta t^2 \Gamma^6 {B'}^2}.
\ee
Now, combining eqs. (1) -- (4), and substituting the values adopted by us
for BL Lacertae, we find
\be
\Gamma \simeq \left ( {L_S^2 \over L_C}\, {\nu_{C}^2 \over \nu_{S}^4}\,
{1 \over \Delta t^2} \,{2 m_e^2 c \over h^2 B_{cr}} \right )^{1/4} \sim
100,
\ee
\be
B' = {1 \over \Gamma} {\nu_{S}^2 \over \nu_{C}} {hB_{cr} \over m_e c^2}
\sim 10^{-4} \; {\rm Gauss},
\ee
and
\be
\gamma_m' = \sqrt {\nu_{C} \over \nu_{S}} \sim 10^5  .
\ee
For such model parameters, the low energy spectral break predicted to arise
due to synchrotron self-absorption should be located at a frequency $\nu_a$
given by
\be
\nu_a \simeq 
1.7 \times 10^2  {B'}^{1/7}(L_{\nu}\nu)_{\nu=\nu_a}^{2/7} \Delta t^{-4/7}
\Gamma^{-5/7} \sim 1.3 \times 10^{10} \; {\rm Hz} ,
\ee
This is lower than the observed value by at least a factor of 10 -- 30
(see, \eg, Bregman \ea 1990).  It should be emphasized here that because no
signature of the high-energy break is observed up to the highest energies
covered by EGRET, the value $h\nu_{C} = 10 $ GeV used here is actually the
lowest plausible value of the location of the high energy break, and that
for a larger $h\nu_{C}$, the output model parameters become even more
extreme.

\subsection{External Radiation-Compton}

The discovery of broad emission lines in BL Lacertae (Vermeulen \ea 1995;
Corbett \ea 1996) indicates that in this object, just as in the case of
quasars, the subparsec jet is embedded in a diffuse radiation field.  The
energy density of this field as measured in the comoving frame of the blob
is thus amplified by a factor $\sim \Gamma^2$.  The rate of the electron
energy losses due to Comptonization of this radiation can then be
approximated by (Sikora \ea 1994)
\be
{d\gamma' \over dt'} \simeq 
{c \sigma_T  u_{ext}' \over m_e c^2} {\gamma'}^2 ,
\ee
where 
\be
u_{ext}' \sim {L_{BEL} \Gamma^2 \over 4\pi r^2 c} ,
\ee
and $r$ is the distance of an active region from the central object (black
hole).  For an opening half-angle of a jet $\theta_j \sim a/r \sim
1/\Gamma$, and for an observer located within or close to the jet cone, the
ratio of the ERC peak luminosity to the synchrotron peak luminosity is
given by (Sikora 1997):
\be
{L_C \over L_S} \simeq
{u_{ext}' \over u_B'} \simeq
{2 L_{BEL} \Gamma^2 \over r^2 c {B'}^2} \simeq 
{2 L_{BEL}  \over c^3 \Delta t^2 {B'}^2 \Gamma^2} ,
\ee
where now
\be
\nu_{C}  \simeq (\gamma_m'\Gamma)^2 \nu_L .
\ee

Combining equations (2), (11), and (12), we obtain
\be
\Gamma \simeq  
\left( 2 L_S L_{BEL} \over L_C \right)^{1/4}
\left( {\nu_{C} \over \nu_{L} \nu_S} {1\over \Delta t} 
{m_e c^{1/2} \over h B_{cr}} \right)^{1/2} \sim 8  ,
\ee
\be
B' = \Gamma {\nu_{S} \nu_{L} \Gamma \over \nu_{C} }  \, {h B_{cr} \over m_e c^2}
\sim 1 \; {\rm Gauss} ,
\ee
and
\be
\gamma_m' \simeq {1 \over \Gamma} \left( \nu_{C} \over \nu_{L} \right)^{1/2} 
 \sim 4 \times 10^3 .
\ee
For these parameters, $\nu_a \simeq 4 \times 10^{11}$ Hz, which is
consistent with the observations (Bregman \ea 1990).  In addition, the
relatively low value of $\Gamma$ is consistent with both the superluminal
expansion data (cf. Mutel \& Phillips 1982) and with the limits derived
from the considerations of the compactness of the source as inferred from
the variability data via opacity to pair production.

\subsection{ERC Radiation plus SSC Radiation}

Provided that both ERC and SSC spectral components are produced in the
Thomson regime and that the observer is located within or near the jet
cone, the ratio of the peak luminosities of these two components can be
approximated by the formula
\be
{L_{SSC} \over L_{ERC}} \simeq
{ u_S' \over u_{ext}' } \simeq 
{1 \over \Gamma^4} {L_S \over L_{BEL}} ,
\ee
Assuming that the $\gamma$--rays are produced by the ERC mechanism, we use
in eq. (12) $\Gamma \simeq 8$ and $\gamma_m' \simeq 4 \times 10^3$ (see \S
3.2) and obtain a luminosity of the SSC radiation of $L_{SSC} \simeq 0.2 \,
L_{ERC} \simeq 0.6 \times 10^{46}$ erg s$^{-1} $ and a location of its peak
at $h\nu_{SSC} \simeq {\gamma_m'}^2 \nu_{S} \simeq 15$ MeV.  The two
spectral components, the SSC and the ERC, overlap for $\Gamma^2 \nu_{L} <
\nu < \nu_{SSC}$.  In this range,
\be
{L_{SSC\nu}  \over L_{ERC\nu}} \simeq
\left( \gamma_{(SSC)}' \over \gamma_{(ERC)}' \right)^{2(1-\alpha_X)} 
{u_S' \over u_{ext}'} \simeq
\left( {h\nu_{L} \over m_ec^2} {B_{cr} \over B'} {\Gamma \over {\gamma_m'}^2}
\right)^{1-\alpha_X} {L_S \over L_{BEL}} 
{1 \over \Gamma^4} \sim 10 ,
\ee
where $ \gamma_{(ERC)}' \simeq \sqrt {\nu / \nu_{L}}/ \Gamma$ and $
\gamma_{(SSC)}' \simeq \sqrt {\nu/\nu_{S}}$ are the energies of the
electrons contributing to the radiation at a frequency $\nu$ via ERC and
SSC process, respectively.

This combined synchrotron + SSC + ERC model predicts that all three
spectral components should have a break due to adiabatic losses of
electrons below a certain energy.  Since radiative energy losses are
dominated by the ERC process, the corresponding break energy of the
electron distribution, $\gamma_b'$, can be found from equating the time
scale of the ERC energy losses, $t_{ERC} \simeq \Gamma \gamma'/
(d\gamma'/dt')_{ERC}$, to the time scale of the propagation of the
perturbed flow pattern, $\Delta t \, \Gamma^2$. Using eqs. (9) and (10), we
obtain
\be
\gamma_b' \simeq {4 \pi m_e c^4 \over \sigma_T }
{\Delta t \Gamma \over L_{BEL}} \sim 10^3 .
\ee
The presence of this break in the electron distribution should result in a
change of slope of $\Delta \alpha \simeq 0.5$ (Sikora \ea 1994) in all
three spectral components.  Note that during a flare, the spectrum most
likely would vary in time as a result of the effective change of the
spectral slope due to the change of the location of this break.  In our
case, we model the time averaged spectrum; in particular, the adiabatic
losses of electrons with $\gamma' < \gamma_b'$ result in the change of the
spectral slope at $\nu ({\gamma_b})$, and the data indeed show $\Delta
\alpha \simeq 0.5$.  In the synchrotron component the break occurs around
0.2 eV, while in the SSC component it is around 0.75 MeV, and in the ERC
spectral component around 0.5 GeV. As calculated from the electron kinetic
equation, this break is very smooth, and since $\gamma_b'$ is very close to
the maximum electron energy, the ``adiabatic'' breaks should join smoothly
with the intrinsic high energy breaks. This is illustrated in Fig. 4, where
the data for the July 1997 flare are fitted with our ERC+SSC model.

As one can see from Fig. 4, the SSC component, calculated self-consistently
within the framework of the ERC model, fits the X--ray data reasonably
well.  We note that such a three-component spectral structure was also
inferred by Kubo \ea (1998) for other QHB blazars.  Our result is that the
high energy ($> 3$ keV) spectrum of BL Lacertae is more similar to blazars
associated with quasars (QHBs) than to HBLs, further strengthening the
inference that LBLs are weak-lined quasars, and that HBLs form a somewhat
distinct subclass of BL Lac type objects.  We also note that in the context
of this model, the overall spectrum of BL Lacertae (and, by similarity,
that of most other LBL-type blazars) is {\sl not} expected to extend to the
TeV energies.  This is because the distribution of the relativistic
electrons does not extend to sufficiently high energies to produce TeV
$\gamma$--rays, while the second-order Comptonization is inefficient
because of the drop in the Compton cross-section due to the Klein-Nishina
limit.  This is in contrast to HBLs, where electron energy distributions
extend to a range that can be $10^2 - 10^3$ times greater (Kubo \ea 1998)
than that derived above.

\section{SUMMARY}

We summarize as follows:  

\vskip 0.1 cm \noindent (1) The RXTE observations of BL Lacertae 
in July 1997, during the flare observed in the ground-based optical and
$\gamma$--ray EGRET data, imply that the source was bright in the X--ray
band.  The X--ray spectrum was relatively hard, exhibiting both flux and
spectral variability with a mean energy power law index $\alpha$ of 0.59.
The source showed a peak in its X--ray flux nearly coincident with the peak
of the GeV $\gamma$--ray flux.

\vskip 0.1 cm \noindent (2) A comparison of the RXTE data to the 
archival {\sl Asca} and ROSAT PSPC data implies that the spectrum of BL
Lacertae is generally harder when the source is brighter.  The {\sl Asca}
data possibly show an additional soft component above the extrapolation of
the hard power law to lower energies (but the presence of this soft
component is uncertain and subject to the details of the Galactic
absorption).  This general spectral behavior appears to be associated with
an emergence of broad emission lines in BL Lacertae first reported in 1995.

\vskip 0.1 cm \noindent (3) The broad-band spectrum of BL Lacertae
appears more similar to a blazar associated with a quasar than to the more
common HBL - type, ``certified lineless'' sub-class of blazars, implying an
association of the presence of the broad emission lines with the high
energy portion of the overall spectrum.  This further supports the
suggestion of Vermeulen \ea (1995) that BL Lacertae is no longer a BL Lac
object; this seems to be true of the high energy portion of its spectrum as
well.

\vskip 0.1 cm \noindent (4) The broad-band spectrum of BL Lacertae 
cannot be readily fitted with either the synchrotron self-Compton (SSC) or
the External Radiation Compton (ERC) model.  However, a hybrid model, where
the X--ray radiation arises via the SSC process and the GeV $\gamma$--ray
radiation is produced via ERC, can fit the data well.  From this model, we
derive the bulk Lorentz factor of the jet $\Gamma \sim 8$, magnetic field
$B \sim 1$ Gauss, and Lorentz factors $\gamma_m'$ of the electrons
radiating in all three components at the peak of their respective $\nu
\times F(\nu)$ distributions to be $\sim 4 \times 10^3$.

\acknowledgments

We acknowledge the referee, Dr. Marscher, for his comments leading to
significant improvements to this paper, and the support from NASA RXTE
observing grants to GSFC via Universities Space Research Association
(USRA), and Polish KBN grant 2P03D00415.

\clearpage

\begin{figure}

\centerline{\bf FIGURE CAPTIONS}

\caption{Light curve obtained from the PCA RXTE observations for 
BL Lacertae in July 1997 over the 3 -- 20 keV bandpass.  The observations
are most likely undersampled, but the highest point in the PCA light curve
roughly corresponding to the highest point in the EGRET light curve,
apparent on 19 July 1997 (cf. Fig. 2 in Bloom \ea 1997).  The systematic
background subtraction errors are $\sim 0.3$ count s$^{-1}$, while the
unknown but constant offset due to the fluctuations of the Cosmic X--ray
Background is $\sim 0.6$ count s$^{-1}$ ($1 \sigma$). }

\caption{Combined July 1997 RXTE PCA and HEXTE data (top
panel) and residuals (bottom panel) to a simple power law model for the
spectrum of BL Lacertae. }

\caption{Unfolded X--ray spectra for BL Lacertae
obtained in December 1992 by ROSAT PSPC, November 1995 by {\sl Asca}, and
July 1997 by RXTE PCA.  These data illustrate the point that the X--ray
spectrum of BL Lacertae appears harder when the source is brighter. }

\end{figure}

\begin{figure}

\caption{Broad-band spectrum of BL Lacertae in July
1997.  {\sc Data from July 1997:} crosses - optical data (Maesano, Massaro,
\& Nesci 1997; Ma \& Barry 1997); solid line ``bowtie'' in the 2 -- 10 keV
band -- RXTE average of the summed July 1997 data (this paper); solid line
``bowtie'' in the 50 -- 300 keV band -- OSSE time averaged data from the
period 15 -- 21 July (Grove \& Johnson 1997); solid line ``bowtie'' in the
$E > 100$ MeV range -- time averaged EGRET data from the period 15 -- 22
July (Hartman \ea 1997; Bloom \ea 1997).  {\sc Data from other epochs}:
dots -- optical, infrared and radio data from various epochs (NED;
V\'eron-Cetty \& V\'eron 1996; Hewitt \& Burbidge 1993); X--rays from {\sl
Ginga} (2 -- 30 keV range from 15 June and 17 July 1988 [Kawai \ea 1991]),
EXOSAT (1 -- 5.6 keV range from 8 June 1980 [Bregman \ea 1980]), and from
{\sl Asca} (dotted line ``bowtie'' in the 2 -- 10 keV range from 22
November 1995 [this paper]); $\gamma$--rays from EGRET (dotted line
``bowtie'' in the $E > 100$ MeV from January 1995 [Catanese \ea 1997]). All
optical data are derredened under assumption $E(B-V)=0.31$ (Bregman 1990)
and using algorithm of Seaton (1979). The long-dashed lines show our
theoretical model, which includes three components: the synchrotron
component dominating in the radio - to - UV range, the synchrotron
self-Compton component in the X-ray range, and the external radiation
Compton component dominating in the MeV - GeV range. }

\end{figure}


\begin{references}

\reference{}
Bania, T., Marscher, A. P., \& Barvainis, R. 1991, \aj, 101, 2147

\reference{}
Blandford, R. D., \& K\"onigl, A. 1979, \apj, 232, 34 

\reference{}
Blandford, R. D., \& Rees, M. J. 1978, in Pittsburgh Conference on BL Lac
Objects, ed. A. M. Wolfe (Pittsburgh University Press), p. 328

\reference{}
Bloom, S. D., \ea 1997, \apj, 490, L145

\reference{}
Bregman, J. N., \ea 1990, \apj, 352, 574

\reference{}
Catanese, M., \ea 1997, \apj, 480, 562

\reference{}
Celotti, A., Padovani, P., \& Ghisellini, G. 1997, \mnras, 286, 415

\reference{}
Corbett, E. A., Robinson, A., Axon, D. J., Hough, J. H., Jeffries, R. D.,
Thurston, M. R., \& Young, S. 1996, \mnras, 281, 737

\reference{}
Crovisier, J., Kazes, I., \& Brillet, J. 1984, \aa, 138, 237

\reference{}
Dermer, C. D., Schlickeiser, R., \& Mastichiadis, A. 1992, \aa, 256, L27

\reference{}
de Vries, H. W., Heithausen, A., \& Thaddeus, P. 1987, \apj, 319, 723

\reference{}
Dickey, J. M., Kulkarni, S. R., van Gorkom, J. H., \& Heiles, C. E. 1993,
ApJS, 53, 591

\reference{}
DuPuy, D., Schmitt, J., McClure, R., van den Bergh, S., \& Racine, R. 1969,
\apj, 156, L135

\reference{}
Ghisellini, G., \& Maraschi, L. 1989, \apj, 340, 181

\reference{}
Ghisellini, G., Maraschi, L., \& Dondi, L. 1996, A\&AS, 120, C503

\reference{}
Giommi, P., Ansari, S. G., \& Micol, A. 1995, A\&AS, 109, 267

\reference{}
Grove, J. E. \& Johnson, W. N. 1997, IAU Circular 6705

\reference{}
Gruber, D., \ea 1996, A\&AS, 120, 641

\reference{}
Hartman, R., \ea 1997, IAU Circular 6703

\reference{}
Hewitt, A. \& Burbidge, G. 1993, ApJS, 87, 451

\reference{}
Jahoda, K., Swank, J. H., Giles, A. B., Stark, M. J., Strohmayer, T.,
Zhang, W., \& Morgan, E. H. 1996, in ``EUV, X--ray, and Gamma-ray
Instrumentation for Astronomy VII'', SPIE Proceedings, O. Siegmund \&
M. Gummin, eds, 2808, 59

\reference{}
Kawai, N., \ea 1991, \apj, 382, 508

\reference{}
Kazes, I., \& Crovisier, J. 1981, \aa, 101, 401

\reference{}
K\"onigl, A. 1981, \apj, 243, 700

\reference{}
Kubo, H., Takahashi, T., Madejski, G., Tashiro, M., Makino, F., Inoue, S.,
\& Takahara, F. 1998, \apj, 504, 693

\reference{}
Ma, F., \& Barry, D. 1997, IAU Circular 6703

\reference{}
Maesano, M., Massaro, E., \& Nesci, R. 1997, IAU Circular 6700

\reference{}
Madejski, G., Jaffe, T. \& Sikora, M. 1997, IAU Circular 6705

\reference{}
Madejski, G., \ea 1997, in ``X--ray Imaging and Spectroscopy of Cosmic Hot
Plasmas,'' F. Makino \& K. Mitsuda, eds. (Tokyo: Universal Academy Press),
229

\reference{}
Maraschi, L., Maccacaro, T., \& Ulrich, M.-H (eds) 1989, ``BL Lac Objects''
(Springer-Verlag: Berlin)

\reference{}
Marshall, F. E., \ea 1980, \apj, 235, 4

\reference{}
Marscher, A. P., Bania, T. M., \& Wang, Z. 1991, \apj, 371, L77

\reference{}
Miller, J. S., \& Hawley S. A. 1977, \apj, 212, L47

\reference{}
Morrison, R., \& McCammon, D. 1983, \apj, 270, 119

\reference{}
Mushotzky, R. F., \& Jahoda, K. 1992, in ``The X--ray Background,''
X. Barcons \& A. Fabian, eds. (Cambridge University Press), 80

\reference{}
Mutel, R., \& Phillips, R. 1982, in ``Extragalactic Radio Sources,''
Proc. IAU Symp. 97, D. Heeschen \& C. M. Wade, eds. (Doerdrecht: Riedel),
385

\reference{}
Noble, J. C., \ea 1997, IAU Circular 6693

\reference{}
Oke, J. B., Neugerbauer, G., \& Becklin, E. E. 1969, \apj, 156, L41

\reference{}
Padovani, P., Perlman, E., Giommi, P., \& Sambruna, R. 1998, BAAS, 30, 1413

\reference{}
Rothschild, R. E., \ea 1998, \apj, 496, 538

\reference{}
Sambruna, R., Maraschi, L., \& Urry, C. M. 1996, \apj, 463, 444

\reference{}
Schmitt, J. 1968, Nature, 218, 663

\reference{}
Seaton, M. J. 1979, \mnras, 187, 73p 

\reference{}
Shafer, R. 1983, PhD thesis, University of Maryland, College Park

\reference{}
Sikora, M. 1997, in ``Proceedings of the Fourth Compton Symposium'' (AIP
Conference Proceedings 410), eds C. D. Dermer, M.S . Strickman \&
J. D. Kurfess, 494

\reference{}
Sikora, M., Begelman, M. C., \& Rees, M. J. 1994, \apj, 421, 153

\reference{}
Strittmatter, P. A., Serkowski, K., Carswell, R. F., Stein, W. A., Merrill,
K. M., \& Burbidge, E. M. 1972, ApJ, 175, L7

\reference{}
Vermeulen, R. C., Ogle, P. M., Tran, H. D., Browne, I. W. A., Cohen, M. H.,
Readhead, A. C. S., \& Taylor, G. B. 1995, ApJ, 452, L5

\reference{}
V\'eron-Cetty, M.-P. \& V\'eron P. 1996, ESO Publication No. 17

\reference{}
Von Montigny, C., \ea 1995, \apj, 440, 525

\reference{}
Wolfe, A. (ed.) 1978, Pittsburgh Conference on BL Lac Objects, (Pittsburgh
University Press)

\reference{}
Worrall, D., \& Wilkes, B. 1990, \apj, 360, 396

\reference{}
Zhang, W., Giles, A. B., Jahoda, K., Swank, J. H., \& Morgan, E. M. 1993,
in ``EUV, X--ray, and Gamma-ray Instrumentation for Astronomy IV,'' SPIE
Proceedings, O. Siegmund ed., 2006, 324

\end{references}
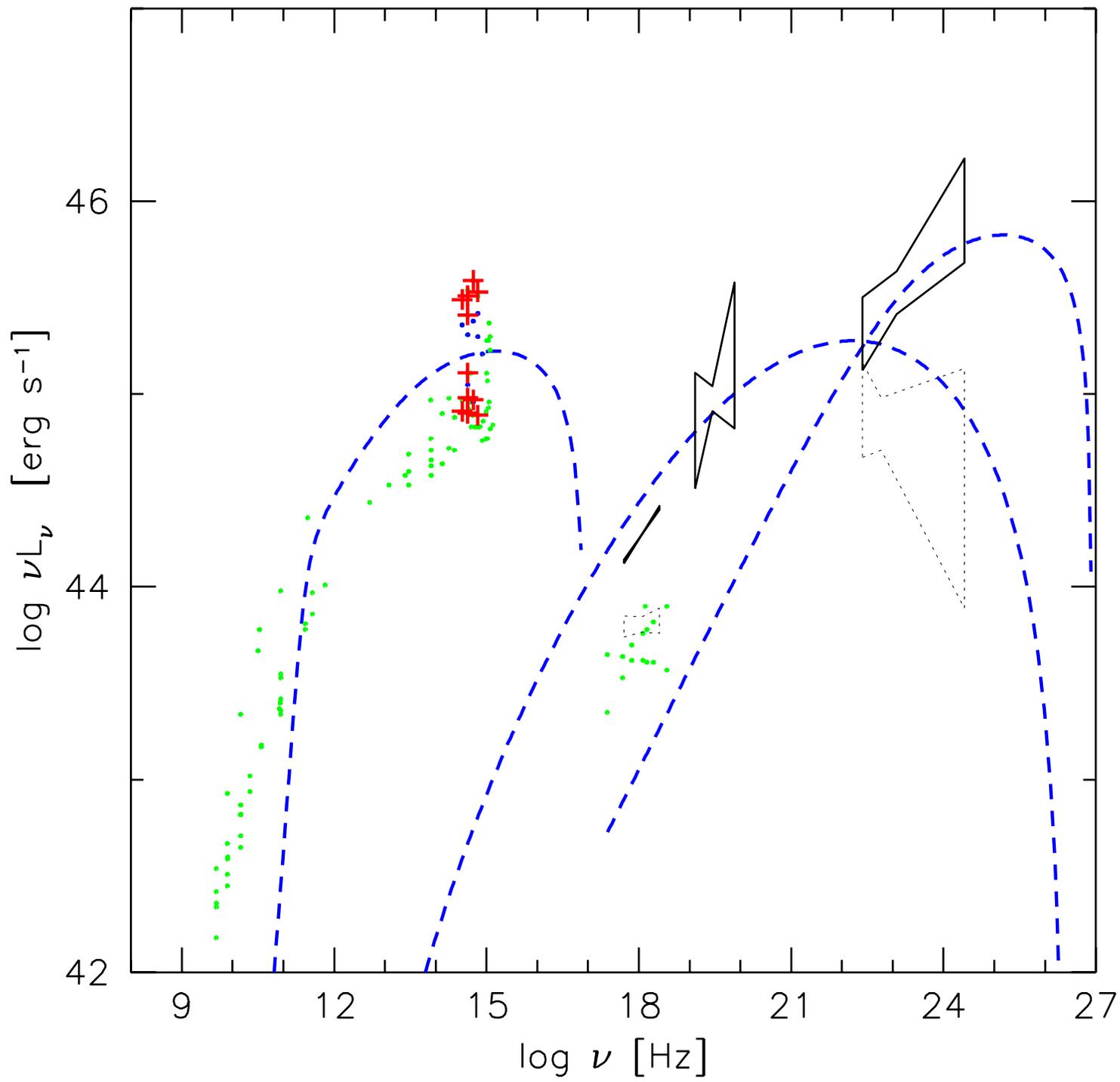
\end{document}

\begin{deluxetable}{ccccccc}
\tablenum{2}
\tablewidth{40pc}
\tablecaption{RXTE Observations of BL Lacertae}
\tablehead{\colhead{Date} & \colhead{Time} & \colhead{Observ.} &
\colhead{Ct. rate} & \colhead{2-10 keV} & \colhead{Power law} &
\colhead{$\chi^2$} \nl
(UT) & (h:m:s) & length (s) & \& error & flux & index \& err & }
\startdata
\multicolumn{7}{l}{\bf PCA:} \nl
16/07/97 & 00:58:56 & 1888 & $5.6 \pm 0.2$ & $2.0 \times 10^{-11}$& $0.60
\pm 0.10$ & 24.9/45 PHA ch. \nl
17/07/97 & 00:57:36 & 1936 & $7.0 \pm 0.2$ & $2.4 \times 10^{-11}$ & $0.57
\pm 0.08$ & 20.7/45 PHA ch. \nl
18/07/97 & 18:48:00 & 1744 & $7.1 \pm 0.2$ & $2.4 \times 10^{-11} $& $0.57
\pm 0.10$ & 30.7/45 PHA ch. \nl
19/07/97 & 07:17:20 & 1728 & $11.0 \pm 0.2$ & $3.7 \times 10^{-11}$ & $0.43
\pm 0.05$ & 36.4/45 PHA ch. \nl
20/07/97 & 13:54:08 & 1936 & $6.9 \pm 0.2$ & $2.5 \times 10^{-11}$& $0.72
\pm 0.09$ & 27.5/45 PHA ch. \nl
20/07/97 & 23:17:52 & 1968 & $9.5 \pm 0.2$ & $3.6 \times 10^{-11}$& $0.81
\pm 0.06$ & 44.1/45 PHA ch. \nl
21/07/97 & 23:15:28 & 1936 & $6.4 \pm 0.2$ & $2.2 \times 10^{-11}$& $0.44
\pm 0.08$ & 26.4/45 PHA ch. \nl

\multicolumn{2}{l}{\bf Full PCA data set} & 13136 & $7.6 \pm 0.05$ & $2.6
\times 10^{-11}$& $0.59 \pm 0.03 $ & 33.4/45 PHA ch. \nl

\tablevspace{10pt}

\multicolumn{7}{l}{\bf HEXTE:} \nl
\multicolumn{2}{l}{\bf Cluster A} & 4138 & $0.76 \pm 0.22$ & & & \nl
\multicolumn{2}{l}{\bf Cluster B} & 4077 & $0.33 \pm 0.19$ & & & \nl

\multicolumn{3}{l}{\bf Combined clusters A+B} & & & $1.2^{+2.0}_{-0.8}$&
136/158 PHA ch. \nl

\tablevspace{10pt}
\multicolumn{3}{l}{\bf Combined PCA + HEXTE} & & & $0.59 \pm 0.03 $ &
172/203 PHA ch.\nl

\tablecomments{ All PCA counting rates are quoted over the 3 -- 20 keV band.
We quote the flux over the 2 -- 10 keV band (in the units of erg cm$^{-2}$
s$^{-1}$) for ease of comparison with previous observations.  All HEXTE
counting rates are over the 20 -- 100 keV band. }

\tablecomments{ Spectral fits to the PCA data were performed using a power
law model over the channels nominally corresponding to the 3 -- 20 keV band
(45 PCA PHA channels), with the absorption fixed at the value measured by
{\sl Asca} ($2.7 \times 10^{21}$ cm$^{-2}$).  In all cases, the absorption
determined from the RXTE data was consistent with that determined from the
{\sl Asca} data; in all cases, the value of the fitted index did not vary
by more than 0.03 when absorption was allowed to be free.  For the HEXTE
data, the fits were performed using channels corresponding to energy range
20 -- 100 keV. }

\enddata
\end{deluxetable}